# Coupled-Band ESSFM for Low-Complexity DBP

Stella Civelli (1,2), Debi Pada Jana(2,3), Enrico Forestieri(2,4), Marco Secondini(2,4)

(1) CNR-IEIIT, stella.civelli@cnr.it
(2) Tecip Institute, Scuola Superiore Sant'Anna
(3) College of Optics and Photonics (CREOL), University of Central Florida
(4) PNTLab, Consorzio nazionale interuniversitario per le telecomunicazioni (CNIT)

***Abstract*** *We propose a novel digital backpropagation (DBP) technique that combines perturbation theory, subband processing, and splitting ratio optimization. We obtain 0.23 dB, 0.47 dB, or 0.91 dB gains w.r.t. dispersion compensation with only 74, 161, or 681 real multiplications/2D-symbol, improving significantly on existing DBP techniques. ©2024 The Author(s)*

## Introduction

Nonlinear propagation effects limit the performance of coherent fiber-optic communications systems[1]–[3]. A possible solution is offered by digital backpropagation (DBP), which can ideally compensate for deterministic intrachannel nonlinear interactions[4],[5].

The implementation of DBP, though possible in theory, is practically limited by the complexity of the available algorithms. Indeed, the naive approach of describing fiber propagation with sufficiently small steps in which linear and nonlinear effects act independently—the split step Fourier method (SSFM)—requires a large computational complexity, making its use for DBP impracticable. Consequently, in the past years, several low-complexity DBP techniques have been proposed[6]. Among these, there are some improved versions of the SSFM, which use a modified nonlinear step to account for the interaction between nonlinearity and dispersion, e.g., the filtered DBP[7],[8] and the enhanced SSFM (ESSFM)[9],[10]. Furthermore, several approaches based on machine learning have been recently proposed[11], either in combination with subband processing[12], or using carrier phase recovery[13].

In this work, we improve the ESSFM by employing subband processing and using asymmetric steps with optimized splitting ratio. The subbands are jointly processed as in the multi-channel DBP algorithm proposed in[14],[15], which allows increasing the step size while still accounting for the cross-phase-modulation (XPM)-like part of intra-band nonlinear interactions. The optimization of the splitting ratio further improves the accuracy of the algorithm by accounting for the asymmetry of long steps due to attenuation. By using numerical simulations, we show that in a 15x80 km link the proposed algorithm needs only 74 real multiplications per 2D symbol (RMs/2D) to achieve a signal-to-noise ratio (SNR) gain of about 0.23 dB with respect to electronic dispersion compensation (EDC). The gain increases to 0.47 dB and 0.91 dB with 161 and 681 RMs/2D, respectively.

## Coupled-Band ESSFM

The coupled-band ESSFM (CB-ESSFM) is a novel technique for single-channel DBP, which combines logarithmic perturbation, subband processing, and splitting ratio optimization to improve the tradeoff between performance and complexity. The CB-ESSFM structure is sketched in Fig. 1. The received channel (in a WDM scenario, after demultiplexing the channel of interest) is digitally demultiplexed into $N_{\text{sb}}$ equally spaced subbands. Next, the subbands are processed by an alternate cascade of $N_{\text{st}} + 1$ linear steps and $N_{\text{st}}$ nonlinear steps, as in an SSFM-like structure with step size $L$. The linear step, applied independently on each band, accounts for group velocity dispersion (GVD) in the frequency domain, as in the conventional SSFM. The nonlinear step processes the subbands jointly. Each subband undergoes a nonlinear phase rotation (NLPR) in time domain that accounts for intra- and inter-band nonlinearity and their interaction with GVD. The NLPRs depend linearly on the intensity of the subbands, from which they are obtained through a MIMO filter, following the same simplified logarithmic-perturbation approach used in the multi-channel DBP algorithm described in[15].[1] The coefficients of the MIMO filter can be optimized numerically or obtained, with good approximation, from perturbation theory. The results in this paper have been obtained by using the first approach. In the typical symmetric

---
[1] The full complexity algorithm proposed in[14] could also be employed. However, since it provides a similar performance, we prefer the reduced-complexity algorithm in[15].

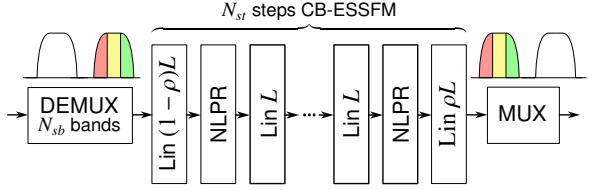

**Fig. 1:** CB-ESSFM processing scheme with $N_{sb} = 3$ bands.

SSFM configuration, each nonlinear step of size $L$ is sandwiched between two half linear steps of size $L/2$. However, when a small number of steps is employed, the resulting configuration is not actually symmetric due to attenuation, which makes the nonlinear interactions in the first half (of the backward link) weaker than those in the second half. Therefore, we also optimize the position of the nonlinear step according to the splitting ratio $0 \leq \rho \leq 1$, such that it is placed between a linear step of size $(1-\rho)L$ and another one of size $\rho L$. Adjacent linear steps can be combined together, so that the overall configuration changes only in the first linear step, of size $(1-\rho)L$, and in the last linear step, of size $\rho L$, while all the remaining $N_{st} - 1$ linear steps have size $L$. Finally, the subbands are digitally multiplexed back together to obtain the backpropagated signal.

The main advantage of the CB-ESSFM is that it allows processing the signal in smaller bandwidths, hence reducing the number of steps required to account for GVD-induced walk-off and to achieve a certain accuracy. However, a naive subband processing would neglect all inter-band interactions, thus strongly hampering the overall performance. By contrast, CB-ESSFM accounts for XPM-like interactions, while neglecting four-wave-mixing-like ones. As a result, CB-ESSFM provides the largest advantages when the number of steps is small, though it cannot achieve exactly the same performance as ideal DBP at a very large number of steps. The optimal number of bands, which depends on the number of steps (when $N_{st}$ is large, one band is optimal), is the one that optimizes the trade-off between intra- and interband accuracy: larger $N_{sb}$ improves intraband compensation (as each band is narrower), smaller $N_{sb}$ improves interband compensation (as less nonlinear interactions are neglected).

The CB-ESSFM can be implemented by means of the overlap and save technique, by processing all signal's samples—defined with $n$ samples/symbols—arranging them in blocks of $N$ samples, with a partial overlap of $N_{ov}$ samples between blocks[16]. For each block, each linear step is implemented in the frequency domain, and each NLPR in the time domain. The whole processing (including subband MUX and DEMUX) requires 4 (direct or inverse) complex FFTs (CFFT) of size $N$, $4N_{sb}N_{st}$ CFFTs of size $N/N_{sb}$, $N_{st} + 1$ GVD compensations, and $N_{st}$ NLPRs (based on frequency-domain MIMO filtering and requiring $2N_{sb}N_{st}$ additional real FFTs of size $N/N_{sb}$). Implementing each complex multiplication with three real multiplications (RMs) and five real additions[17], and using the split-radix algorithm[18] to implement the FFTs, the overall number of RMs/2D required by the CB-ESSFM algorithm is

$$C_{\text{RM}} = \frac{n}{2} \frac{N}{N - N_{ov}} \left( (5N_{st} + 4) \log_2 \frac{N}{N_{sb}} + \right.$$
$$+ N_{st} \frac{3N_{sb} + 1}{2} + 4 \log_2 N_{sb} - 6 + \quad (1)$$
$$\left. + \frac{20 N_{sb} N_{st} + 16}{N} \right) \quad \text{RMs/2D symb.}$$

**System setup and performance**

We test the performance of the proposed DBP technique by means of simulations. The transmitted WDM signal is composed of 5 channels, each with baud rate $R_s = 93$ GBd and $100$ GHz spacing, and uses a dual-polarization uniform $64$ quadrature amplitude modulated (QAM) constellation to modulate a root-raised-cosine pulse with rolloff $r = 0.05$. The link consists of $15$ spans of $80$ km single mode fiber (attenuation $\alpha_{\text{dB}} = 0.2$ dB/km, dispersion $D = 17$ ps/nm/km, and nonlinear parameter $\gamma = 1.27$ W$^{-1}$km$^{-1}$). The loss is compensated after each span by an erbium-doped fiber amplifier with a noise figure of 4.5 dB. The receiver demultiplexes the central channel and applies either EDC or DBP with $n = 1.125$ samples/symbol and block length $N = 16384$. Finally, matched filtering, sampling at symbol time, and mean phase rotation removal are applied. Performance is shown in terms of received SNR at optimal launch power.

Fig. 2 shows the performance of CB-ESSFM with 1 band (dashed)—equivalent to ESSFM when $\rho = 0.5$—and 2 subbands (solid) as a function of the splitting ratio $\rho$ for $N_{st} = 1, 15, 30$. In all considered cases, the use of subband processing with two bands turns out to be advantageous with respect to single-band ESSFM. In fact, though not shown here, the optimal number of bands is 2 when $N_{st} \leq 30$. Next, the figure shows that when a single-step DBP is considered ($N_{st} = 1$) the symmetric configuration (with $\rho = 0.5$) is optimal. Conversely, for 15 or 30 steps, besides an obvious increase of the overall SNR, the figure shows that using asymmetric configurations with smaller splitting ratios can improve the performance, both with $1$ or $2$ subbands. In particular, with respect to the symmetric case with $N_{\text{sb}} = 2$, a gain of $0.26$ dB is obtained with $\rho = 0.12$ for $N_{\text{st}} = 15$, and a gain of $0.1$ dB is obtained with $\rho = 0.23$ for $N_{\text{st}} = 30$. We conjecture that the optimal configuration is obtained when each step is divided into two portions with similar accumulated nonlinear interactions. Due to fiber attenuation, this is obtained for $\rho \approx 0.5$ for $N_{\text{st}} \ll N_{\text{sp}}$ (e.g., $N_{\text{st}} = 1$ in Fig. 2) or for $N_{\text{st}} \gg N_{\text{sp}}$ (not shown here since too complex for a practical implementation), but possibly for smaller $\rho$ when $N_{\text{st}}$ and $N_{\text{sp}}$ are comparable. For instance, with $N_{\text{st}} = 15$ there is exactly one step per span, so that most nonlinear interactions take place in the first portion of each (forward) span, corresponding to the last portion of each DBP step. Thus, a better balance is obtained with a small $\rho$. A similar reasoning can be done for $N_{\text{st}} = 30$, which uses exactly $2$ steps per span. In this case, the impact of attenuation on each step is smaller than for $N_{\text{st}} = 15$ (but still significant). This results in a higher optimal splitting ratio (but still lower than 0.5). In the following, we will always consider the CB-ESSFM with optimized splitting ratio and $N_{\text{sb}} = 2$.

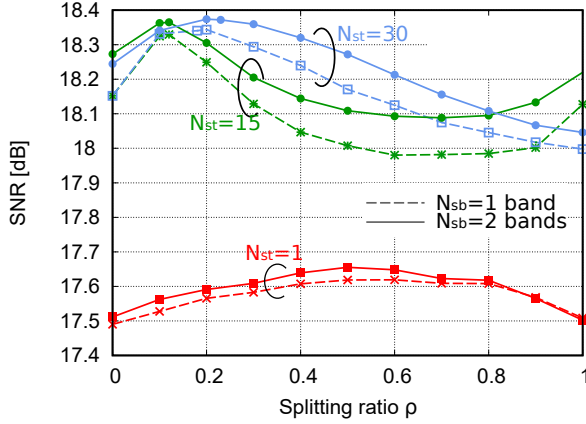

**Fig. 2:** SNR versus splitting ratio with CB-ESSFM with $1$ and $2$ bands for different number of steps $N_{\text{st}}$.

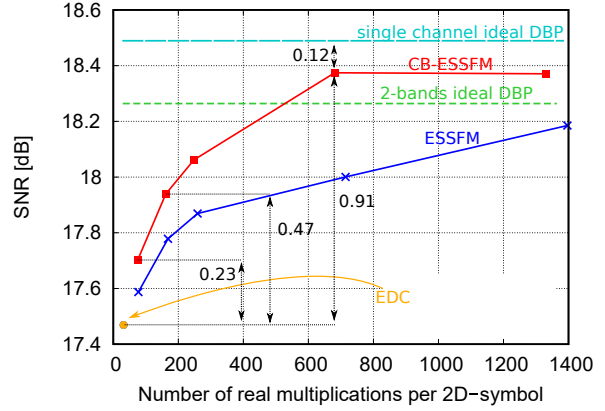

**Fig. 3:** SNR versus complexity for EDC or different DBP tecniques.

Fig. 3 compares the performance of different DBP techniques and EDC as a function of the complexity (1). The single-channel and 2-band ideal DBP curves show, respectively, the maximum performance achievable by applying ideal DBP to the whole channel or by dividing it into two subbands and applying ideal DBP separately to each of them. While the former provides perfect intrachannel nonlinearity compensation, the latter neglects nonlinear interactions between the two subbands. The two limits are numerically obtained by considering the conventional SSFM algorithm and increasing the number of steps until the performance saturates to its maximum value. The corresponding complexity exceeds by more than one order of magnitude the range considered in the figure and is totally unfeasible for a practical implementation. The CB-ESSFM is shown here for the optimal number of bands $N_b = 2$, and with optimal splitting ratio $\rho$, while ESSFM is shown in the conventional symmetric configuration ($\rho = 0.5$). Overall, the figure shows that CB-ESSFM outperforms ESSFM and saturates to its maximum performance with just 681 RMs/2D (corresponding to $N_{\text{st}} = 15$), achieving a gain of 0.91 dB compared to EDC—better than 2-band ideal DBP and close to single-band ideal DBP, with a gap of only $0.12$ dB. This small gap is due to the fact that CB-ESSFM can compensate exactly for intraband nonlinearity, but only approximately for interband nonlinearity. Moreover, CB-ESSFM provides significant gains even at much lower complexity, e.g., 0.23 dB at 74 RMs/2D ($N_{\text{st}} = 1$), 0.47 dB at 161 RMs/2D ($N_{\text{st}} = 3$), and 0.6 dB at 248 RMs/2D ($N_{\text{st}} = 5$).

**Conclusion**

We have proposed a novel method for single-channel DBP, improving our previously proposed ESSFM by means of subband processing and by optimizing the splitting ratio of the propagation steps. Subband processing allows reducing the number of steps (hence the complexity) required to achieve a desired performance, while the optimization of the splitting ratio improves the performance with a given number of steps. In a 15x80 km link with 5 WDM channels, by properly selecting the number of steps, we obtain effective SNR gains ranging from 0.23 dB to 0.91 dB compared to EDC, with a corresponding complexity ranging from 74 to 681 real multiplications per 2D-symbol.

**Acknowledgments**

This work was partially supported by the European Union under the Italian National Recovery and Resilience Plan (NRRP) of NextGenerationEU, partnership on "Telecommunications of the Future" (PE00000001 - program "RESTART").